\documentclass[sn-aps, iicol, Numbered]{sn-jnl}

\usepackage{graphicx}%
\usepackage{multirow}%
\usepackage{amsmath,amssymb,amsfonts}%
\usepackage{amsthm}%
\usepackage{mathrsfs}%
\usepackage[title]{appendix}%
\usepackage{xcolor}%
\usepackage{textcomp}%
\usepackage{manyfoot}%
\usepackage{booktabs}
\usepackage{algorithm}%
\usepackage{algorithmicx}%
\usepackage{algpseudocode}%
\usepackage{listings}%
\usepackage{tikz}
\usepackage{subcaption}

\DeclareCaptionLabelSeparator{none}{ }
\begin{document}

\title{Influence based explainability of brain tumors segmentation in multimodal Magnetic Resonance Imaging}


\author[1,2]{\fnm{Tommaso} \sur{Torda}}

\author[1,2]{\fnm{Andrea} \sur{Ciardiello}}

\author[1,2]{\fnm{Simona} \sur{Gargiulo}}

\author[2]{\fnm{Greta} \sur{Grillo}}

\author[1,2]{\fnm{Simone} \sur{Scardapane}}

\author[1,2]{\fnm{Cecilia} \sur{Voena}}
\equalcont{Corresponding author: Cecilia Voena, cecilia.voena@uniroma1.it}

\author[1,2]{\fnm{Stefano} \sur{Giagu}}

\affil[1]{\orgname{``Sapienza" University of Rome}, \orgaddress{\street{Piazzale Aldo Moro, 5}, \city{Rome}, \postcode{00185}, \state{Italy}, \country{UE}}}

\affil[2]{\orgname{INFN Sezione di Roma}, \orgaddress{\street{Piazzale Aldo Moro, 5}, \city{Rome}, \postcode{00185},   \state{Italy}, \country{UE}}}


\abstract{
In recent years Artificial Intelligence has emerged as a fundamental tool in medical applications. Despite this rapid development, deep neural networks remain black boxes that are difficult to explain, and this represents a major limitation for their use in clinical practice. We focus on the segmentation of medical images task, where most explainability methods proposed so far provide a visual explanation in terms of an input saliency map. The aim of this work is to extend, implement and test instead an influence-based explainability algorithm, TracIn, proposed originally for classification tasks, in a challenging clinical problem, i.e., multiclass segmentation of tumor brains in multimodal Magnetic Resonance Imaging. We verify the faithfulness of the proposed algorithm linking the similarities of the latent representation of the network to the TracIn output. We further test the capacity of the algorithm to provide local and global explanations, and we suggest that it can be adopted as a tool to select the most relevant features used in the decision process. The method is generalizable for all semantic segmentation tasks where classes are mutually exclusive, which is the standard framework in these cases.  }

\keywords{Artificial Intelligence, Explainability, Deep Learning, Healthcare,  Brain Tumors, Segmentation}

\maketitle
\section{Introduction}
The implementation of Artificial Intelligence (AI) algorithms in the medical domain is continuously increasing, driven by advances in the AI field both from the algorithm and the computational power side, particularly in medical image analysis. AI models can perform several tasks on medical images nowadays \cite{Celard2023,Qureshi2023,Jiang2023,Kaur2023}, and among these is the segmentation task, which is essential for diagnosis, treatment planning and monitoring in the clinical management of diseases. 

Medical image segmentation consists of the outline of an organ or a lesion in a medical image and it is often performed by a clinician (radiologist). Deep Neural Networks (DNN), in particular Convolutional Neural Networks (CNN) \cite{Girshick2014} have proved their effectiveness in aiding in this task.

On the other hand, the adoption of AI solutions in clinical practice is still strongly limited by a lack of trustworthiness due to the little transparency of decisional processes and validation mechanisms of such complex models \cite{Hassija2024}. Moreover, from a legal point of view, an innovative aspect of the General Data Protection \& Regulation (GDPR),  adopted by the European Community, is the right of explanation for all individuals to obtain a meaningful explanation of the logic involved where an automated decision is taken. Interpretation of decisions in some level of detail to ensure that algorithms perform as expected and to increase the trust in these methods by the final users represent huge open scientific challenges, not only in the medical field \cite{Arrieta2019,Saeed2023,Gerlings2021}. This has increased the level of interest in developing and testing methodologies that allow the interpretation of AI algorithms' predictions in terms of transparency and explainability, which has become today one of the most important open questions in AI. From now on, we refer to this AI field as to explainable AI (xAI). It should be stressed that xAI can be also an important tool to cross-check DNN models beyond standard performance metrics (e.g., accuracy) and to identify possible errors and biases before deploying a tool into clinical practice \cite{Weber2023,Markus2021}.

In the medical image analysis domain, there is a wide literature about xAI methods with most work related to classification tasks. We refer to \cite{Velden2022,Borys2023,Hassija2024} for recent and comprehensive reviews. 

A useful taxonomy divides algorithms based on their scope into global and local methods \cite{Hassija2024}:
global interpretability approaches aim to enhance understanding of a model's overall logic and the justification behind specific predictions.
On the other hand, local interpretability focuses on providing explanations for each individual choice and prediction rather than offering a detailed description of the complex mechanism underlying the entire black-box model. 
Among local methods, one of the most commonly used techniques is feature attribution. Saliency map methods, a subset of feature attribution-based methods, formulate their explanation based on the relative importance of all input features. This information is usually provided visually by highlighting image regions that are crucial for the model's predictions.
The saliency method explains the algorithm's decision by assigning values that reflect the importance of input components in contributing to that decision.

Many approaches applied to xAI in medical image segmentation provide this type of visualisation of the relevant information in the intermediate layers of the DNN, for example using Grad Class Activation Mapping (CAM) \cite{Natekar2020} or Guided Back-Propagation techniques \cite{Wickstroem2020} to explain 2D brain tumor segmentation of Magnetic Resonance Imaging (MRI) images and semantic segmentation of colorectal polyps from colonoscopy images, respectively. In an alternative approach \cite{Pereira2018}
a set of features obtained from an ML system is used to provide xAI in a brain tumor DNN. In \cite{Saleem2021}, a CAM-based approach is extended to generate a visual explanation for a 3D DNN segmenting brain tumor, using a gradient-independent methodology.

Saliency maps have been criticized for being visually appealing but misleading, noisy, and lacking sensitivity to both the model and the data utilized \cite{Adebayo2020}. Additionally, in brain multimodal segmentation, they have been found inadequate in reflecting modality-specific feature importance \cite{Jin2021}. Moreover, since segmentation involves localization, interpreting saliency maps in segmentation tasks can be inherently ambiguous. Highlighted regions may correspond to both the segmented object and the surrounding context \cite{Arun2021}. Moreover, saliency maps are fundamentally redundant once the segmentation map is known.

 An open issue that needs to be addressed is how to evaluate the quality of an explanation \cite{Nauta2023}. Various figures of merit have been identified, such as the ease of use, the plausibility (correctness of the explanation and correspondence to what the user expects), the faithfulness (how accurately the explanation reflects the model's true decision process), and the robustness (effect on the explanation of changing some aspects of the DNN model or training process), but a standard does not exist yet. Moreover, the explanation must be helpful for end users, in this case, radiologists and clinicians, making this problem interdisciplinary \cite{Loetsch2022}. 
 
 For this reason, it becomes even more important to test the faithfulness of the xAI algorithm with respect to the network's decision-making model. Indeed, it is well known that end-users tend to assume that the explanation is faithful to the decision model \cite{Jin2022}, leading them to underestimate the importance of a faithful explanation and instead favour the accuracy of the model to place trust in the neural network \cite{Papenmeier2019}. It is therefore increasingly important not only to propose new $post$-$hoc$ explanation methods but also to benchmark \cite{Hooker2019} their faithfulness using shared criteria \cite{Yin2022}, such as the Removal-based Criterion based on the idea that the presence of a set of important features identified by a faithful interpretation should have a more meaningful influence on the model's decision than an arbitrary set of features. By measuring how much the model performance would drop after the set of the ``most relevant'' features in an explanation is removed, we can quantify the faithfulness of the xAI algorithm.

\section{Contributions of this work}
In this paper we are interested in extending and applying in the medical context a different class of xAI methods, known as example-based explanations, that do not suffer from the same limitations of feature attribution methods. These methods justify a prediction of black-box models using specific instances from the model's training dataset.
Machine learning methods generate predictions based on the inferences made from their training data. Even a slight modification in a training instance could significantly alter the resulting model.
An influential instance refers to a data instance whose removal strongly impacts the trained model. The more the model parameters or predictions change when the model is retrained with a particular instance removed from the training data, the more influential that instance is.

An interesting xAI algorithm that has become very popular over the past years is TracIn \cite{Pruthi2020}. 

The main idea behind the algorithm is that two similar examples in the dataset, and so influential for a prediction, during training would be responsible for similar parameter updates in the model. 

TracIn was originally developed for classification tasks  \cite{Pruthi2020} and to the best of our knowledge has never been used in segmentation tasks so far. The output of the algorithm applied to the case under study is a score based on which the examples of the training set can be divided into proponents, which have a positive influence on the prediction since they served to reduce loss, opponents, which have a negative influence on the prediction, and neutrals, that have no influence.

In this paper, we want to extend the TracIn algorithm to the task of brain tumor multiclass segmentation in multimodal MRI.  We chose this task based on the availability of public annotated datasets and AI models with good performances. In fact, brain segmentation is one of the first tasks where neural networks have been successfully implemented and many different models have already been tested and validated. Moreover, the clinical case is quite relevant, since brain tumors are one of the leading causes of death worldwide \cite{Wadhwa2019}.

We consider as reference dataset BraTS19 \footnote{https://www.med.upenn.edu/cbica/brats2019/data.html}, from the BraTS tumor segmentation challenge. MRI scans are widely used for the clinical management of this disease, since they allow to capture of different properties of the tumor itself, by using various modalities. The most commonly used modalities are T1-weighted (T1w), T2-weighted (T2w), post-contrast T1-weighted (T1Gd) and T2 Fluid Attenuation Inversion Recovery (T2-FLAIR). 
We use a standard 2D UNet \cite{Ronneberger2015} for the segmentation task.

The final goal is to allow the end user, the radiologist, to visualize the best proponents of the patient under consideration so that he can gain trust in the DNN segmentation by checking the coherence of the decision with the medical domain knowledge (local explanation).

To this purpose, we propose an extension of the original TracIn model to a multiclass segmentation task. In the proposed extension, we decorrelate the signal of pixels of different classes, while standard TracIn would implement an average over all pixels. We show that our modification makes it easier to identify unambiguous features between proponents, opponents and neutral influences in a segmentation scenario. 


In addition to a methodology for applying TracIn to the case of multiclass segmentation, we also try to provide an interpretative map of the results, based on the similarity between the features extracted from the convolutional kernels of the UNet of different patients and their proponents/opponents. This would also prove the faithfulness of the xAI algorithm with respect to the model decision (global explanation), in analogy with \cite{Jin2022} and can be used to select the most relevant features of the model.

\section{Material and Methods} \label{MAtMet}

\subsection{Image database} 
\label{subsec:Brats2019Dataset}

BraTS \cite{Menze2015,Bakas2017,Bakas2019} is a brain tumor segmentation challenge aimed at evaluating state-of-the-art methods for the segmentation of brain tumors in multimodal MRI. The 2019 edition focused on the segmentation of gliomas, an intrinsically heterogeneous brain tumor.

\begin{figure*}[!h]
\centering
  \includegraphics[width=\linewidth, scale=0.4]{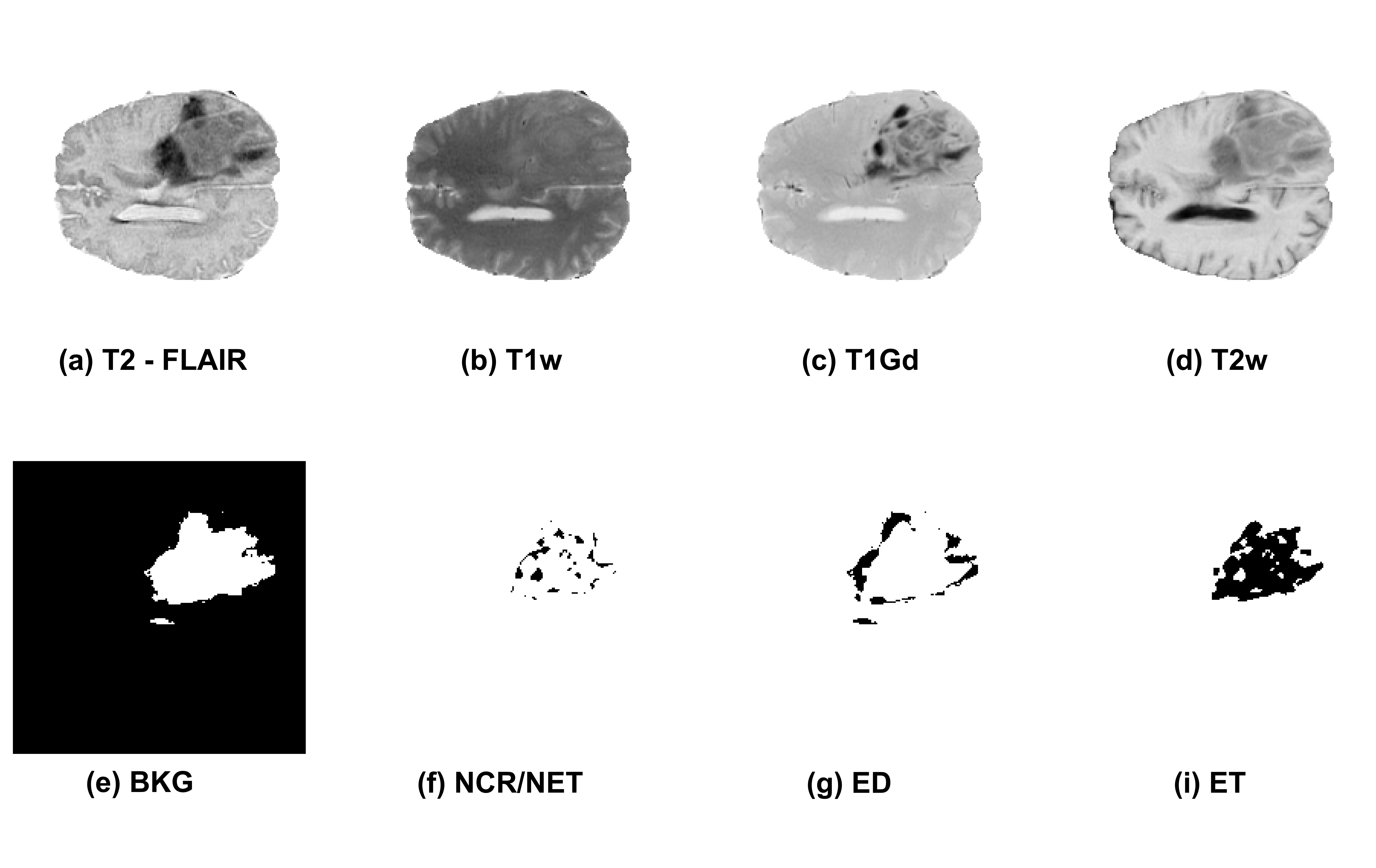}
 \caption{Example of a patient from the BraTS19 dataset. The upper row of images shows a slice of an MRI scan in the four acquisition modalities, the bottom row shows the segmented regions in the four classes}
\label{fig:brats1}
\end{figure*}

BraTS2019 utilizes multi-institutional annotated pre-operative MRI scans. 
The training dataset is composed of $259$ cases of high-grade
gliomas (HGG) and $76$ cases of low-grade gliomas (LGG), manually annotated by both clinicians and board-certified radiologists. For each patient, four MRI scans taken with different modalities are provided: T1, T1Gd, T2, and T2-FLAIR with an image's shape of voxels $240 \times 240 \times 155$. Each MRI has $155$ slices in the sagittal direction. In the manual label of BraTS19 four classes are provided: the GD-enhancing tumor (ET - label 4), the peritumoral edema (ED - label 2), the necrotic and non-enhancing tumor core (NCR/NET - label 1) and the background (BKG - label 0). In the following, we will refer to the GD-enhancing tumor (ET) as label 3 instead of label 4. The ET is described by hyper-intense areas in T1Gd compared to T1 and healthy areas. The appearance of NCR/NET is typically hypo-intense in T1Gd when compared to T1. The peritumoral  ED is typically depicted by a hyper-intense signal in T2-FLAIR.
Figure ~\ref{fig:brats1} shows an MRI of a patient in the four modalities and the segmented regions for the four labels.

We focused only on HGG patients, dividing the dataset into a training set with $207$ patients and a validation set with $52$ patients.  

\subsection{AI algorithm for the segmentation task}
We use a popular and well-established neural network, the UNet \cite{Ronneberger2015} since it performs well, is easy to implement and has extensive literature. One of the reasons why UNet became the state-of-the-art for the segmentation task in medical image analysis is the use of skip connections between the layers of the architecture, which helps the neural network to recognize not only the local features like a classical CNN, but also the global features.\\
We implement a 2-dimensional UNet and we consider separately each slice along the sagittal direction. The input is the MRI image in the four modalities, used as independent ``color'' channels, while at the output stage, each pixel is assigned to one of the four mutually exclusive classes, represented by the tissue type: 0 (BKG), 1 (NCR/NET), 2 (ED), 3 (ET). A predicted mask is built for each class. We select only the central volume of the brain, thus reducing the number of input slices to 10 for each patient, since they are sufficient to cover the tumor and because influence xAI methods are quite computationally expensive. We also crop the image in the \textit{x-y} axes from $240$ to $192$ pixels and we apply elastic transformations, random crop and mirroring, with a probability of 50\% in the data augmentation step. We finally normalize the intensity between $[-1,1]$ and use the softmax function to normalize the output logits. 

We choose the Dice coefficient to both define the loss function and evaluate the algorithm's performance. Given the predicted mask ($P_i$) for the class $i$ and the ground truth (GT$_i$) segmented by the clinician, the Dice coefficient $D_i$ is defined as:
\begin{equation}
    D_i = \frac{2 (P_i \cap \text{GT}_i)}{P_i \cup \text{GT}_i}.
\end{equation}
This metric measures the overlap between the two regions and it is used when the region of interest is smaller than the background area. 

The loss function is defined as (class $0$ is excluded):
\begin{equation}
    \mathcal{L} = 1 - \frac{1}{3}\sum_{i \in [1,3]} D_i.
\label{eq:dice}
\end{equation}
We perform the training of the network using the Adam optimizer with a learning rate of 1e-4 and weight decay of 1e-5 for the first 5 epochs. We switch to stochastic gradient descent (SGD) for the remaining epochs. In total, we train the model for 30 epochs, with the train batch size equal to 10.

Results of the training process are reported in Table ~\ref{tab:score}. 
The performances obtained, expressed using the final Dice score for each class, are in line with those presented by the BraTS2019 challenge.

\begin{table}[!h]
    \centering
    \begin{tabular}{c|c|c|c}
     \hline
         & NCR/NET & ED & ET \\  \hline
         Training set & 0.90 & 0.93 & 0.93 \\
        Validation set & 0.77 & 0.87 & 0.87 \\
          \hline
    \end{tabular}
    \caption{Dice score obtained by the UNet on the BraTS19 dataset (training set and validation set) }
    \label{tab:score}
\end{table}

\subsection{xAI algorithm: extended TracIn} \label{subsec:xAIMethods}
The idea of influence-based xAI methods is to estimate the effect of each example $z$ of the training set $\mathcal{D}$, composed of $n$ elements, on a particular network prediction on an instance $z'$, usually from an independent test set. This can in principle be done by removing one train example at a time and by retraining the network, which is quite computationally expensive. The influence $I(z,z')$ can be defined as:
\begin{equation}\label{eq:theoretical_inf}
    I(z,z') = \mathcal{L}(z'; \theta; \mathcal{D}) - \mathcal{L}(z'; \bar{\theta}; \bar{\mathcal{D}}),
\end{equation}
where  $\mathcal{L}$ is the loss, $\mathcal{D}$ ($\bar{\mathcal{D}}$) and $\theta$ ($\bar{\theta}$), the dataset and the parameter set including and excluding $z$.

Pruthi et al. \cite{Pruthi2020} implemented an approximation of the influence that monitors loss changes during the training when a training example is used to make a prediction, in a first-order gradient approximation. Given each pair ($z,z'$), the $\text{TracIn}(z,z')$ score is expressed as:
\begin{equation}
    \text{TracIn}(z,z') = \sum_{t \in \mathcal{C}} \eta_t \nabla \mathcal{L}(\theta_t \;, z') \cdot \nabla \mathcal{L} (\theta_t \;, z),
    \label{eq:tracin}
\end{equation}
where $\mathcal{C}$ are the checkpoints where the computation is performed, $\theta_t$ the parameters and $\eta_t$ the learning rate at the checkpoint $t$. The definition has been obtained independently from architecture, domain and task, but the use of stochastic gradient descent (SGD)  is assumed. The choice of checkpoints is relevant: in later checkpoints the loss flattens assuming the training is converging and loss gradients are small, earlier checkpoints are usually more informative.

Based on the TracIn score, $z$ is called proponent (opponent) of $z'$ if it has a positive (negative) value of the influence score. Neutral train examples have TracIn scores close to zero. By sorting train examples in order of descending TracIn score, we can identify the strongest proponents (highest score) and strongest opponents (lowest score).

With respect to the original work \cite{Pruthi2020} and subsequent applications, which considered classification tasks, our use-case has the complication of being a segmentation task, which can be viewed as several simultaneous classifications (one for each pixel), and an extension is necessary.

With the loss defined in Equation \eqref{eq:dice}, Equation \eqref{eq:tracin} becomes:
\begin{equation}
    \text{TracIn}(z,z') = \sum_{t \in \mathcal{C}} \eta_t \sum_{(i,j) \in \text{Cl} }\nabla D_j(z) \cdot \nabla D_i(z'),
    \label{eq:tracin2}
\end{equation}
where Cl refers to the set of segmentation classes. In our case, we exclude the background class and we consider the others (NCR/NET class 1, ED class 2, ET class 3), while $z$ and $z'$ are MRI slices. From now on, we refer to $z$ and $z'$ as a train and a test image, respectively.

It should be noted that, given the expression of the Dice score, $D_j$($z$) ($D_i$($z'$)) is computed on the region of $z$ ($z'$) predicted to be of class $j$ ($i$). Equation  \eqref{eq:tracin2} thus mixes influence contributions of quite heterogeneous pixels, introducing noise in the algorithm. To limit this effect, we propose an alternative definition. First, we split $z$ and $z'$ in separate regions, according to the network prediction: $z_j$ and $z_i'$ are the regions predicted to be of class $j$ and $i$ in $z$ and $z'$, respectively. 

We then define a TracIn score for each pair 
($z_j,z_i'$):
\begin{equation}
    \widetilde{T}_{i,j} (z_j,z'_i)  = \sum_{t \in \mathcal{C}} \eta_t \nabla D_j(z_j) \cdot \nabla D_i(z'_i).
    \label{eq:tracin4}
\end{equation}
The diagonal terms ($i$ = $j$) represent the influence that the region of $z$, predicted to be of class $i$, has on the predicted region of the same class of $z'$, while off-diagonal terms quantify the influence of regions of $z$ predicted to be of classes different from $i$.
Given a test image $z'$ to be explained, for each region predicted to be of class $i$, our algorithm returns an explanation vector $T(z_i')$, composed of the scores $\widetilde{T}_{i,j}$, computed for all the train images $z$ and classes $j$. The explanation vector has then dimension $n \times \text{Cl} $, for each region $i$.
In our definition, each training image $z$ contributes with three scores, one for each class $j$.

We consider as checkpoints the first 5 epochs after the transition to SGD and to be more robust against prediction errors we consider pixels to belong to a region in the associated segmentation map only if the network output is greater than a predefined threshold, fixed to 0.8. This selection removes mainly pixels belonging to the borders of the predicted region.

A prerequisite for the algorithm's definition is that the segmentation classes must be mutually exclusive. However, this is not a strong constraint because it is the standard framework for this type of task. In the case of a single class (tumor - no tumor) Equation \eqref{eq:tracin4} becomes simpler since it contains only diagonal terms.

\begin{figure*}[!h]
\centering
  \includegraphics[width=\linewidth, scale=0.6]{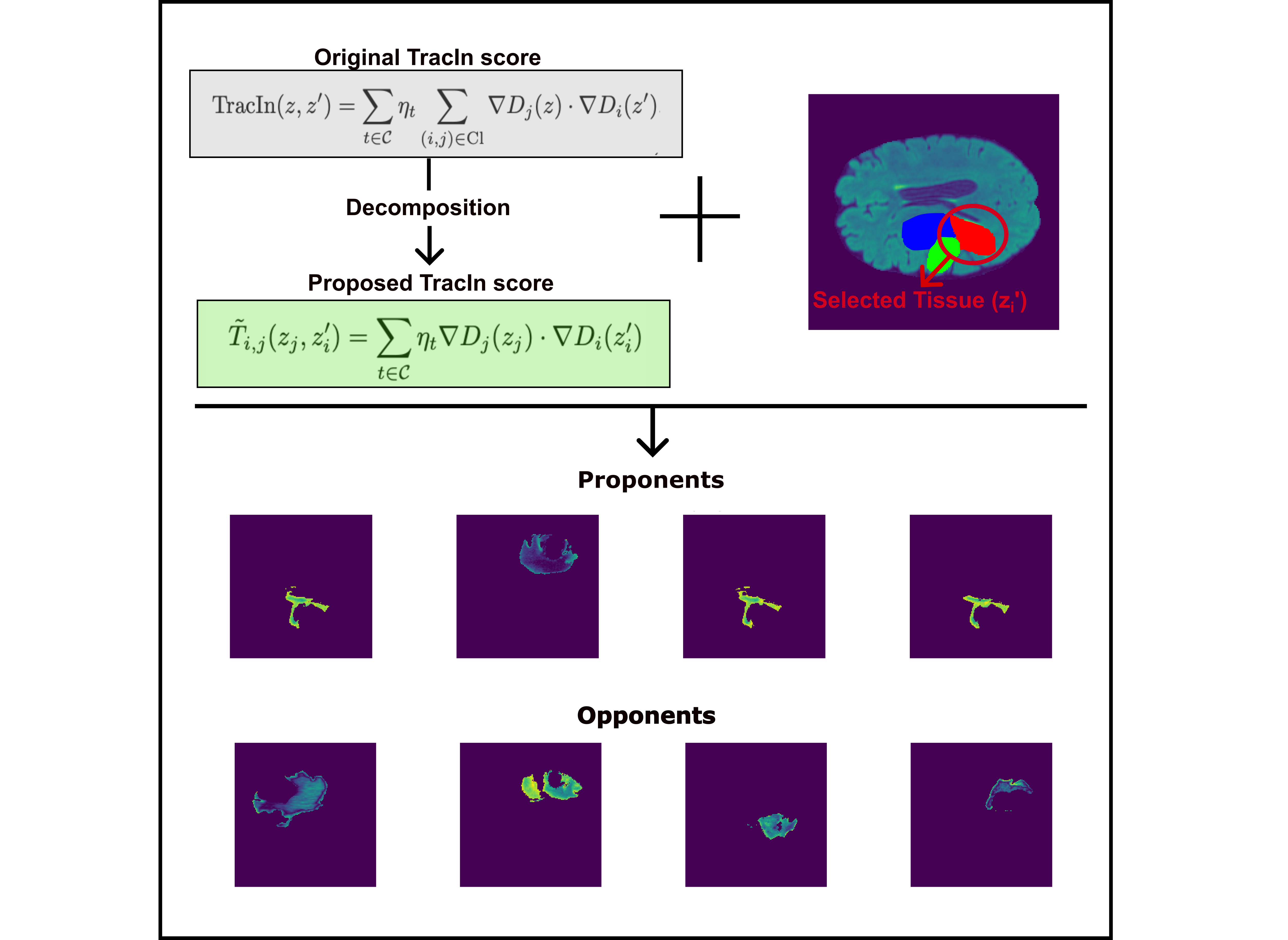}
 \caption{Scheme of the proposed methodology for TracIn calculation in the multiclass segmentation task considered. TracIn is computed separately for each region of the test image $z'$ predicted to be of class $i$ ($z'_i$). Each train image $z$ is also split into regions according to the network prediction ($z'_j$). A threshold on the network output is applied to eliminate pixels with greater prediction errors. For a given $z'_i$ a list of proponents and opponents is provided}
\label{fig:met}
\end{figure*}

Figure \ref{fig:met} shows a scheme of the proposed methodology.

\subsection{Algorithm characterization}
\label{subsec:self}
After defining a methodology for calculating TracIn for segmentation tasks, we carry out studies to understand the algorithm behaviour and to verify the consistency of the definition.
As discussed in \cite{Pruthi2020} an important evaluation technique for this kind of xAI algorithm is the analysis of the train-influence, i.e. the influence of the train set on itself. For example, it is expected that outliers, noisy, corrupted or otherwise difficult to learn objects have a strong influence on themselves (self-influence). 
With our definition, we have three train-influence matrices, one for each class $i$. The matrices dimension is $n \times \text{Cl} \cdot n$.

\subsection{Interpretation maps and faithfulness metric}\label{subsec:Int}
It is well known that post-hoc explainability methods can lead to important misunderstandings about the explanation of the model~\cite{Kaur2023}. It has also been pointed out \cite{Jin2022} how end users tend to take the faithfulness of the explanation method to the network decision model for granted.
In the case of explanations based on visual techniques, visualization of the explanation could be not sufficient to increase the trust in the network prediction. Indeed, the interpretation of it risks being conveyed by the subjective biases of the user, who finds himself judging the goodness of the explanation without having much useful information at his disposal. In the case of TracIn, for example, we can observe which train examples helped or not the prediction (proponents and opponents), but we have no clue as to why this choice was made. 

To overcome this problem, we propose a method for evaluating the faithfulness of TracIn with respect to feature maps extracted from the network. The goal is to find a relation between the latent representation in the network of $z$ and $z'$, and the TracIn score associated with the pair ($z$,$z'$). 
Assuming we have a train set $D_{\text{train}}$, a test set $D_{\text{test}}$ and a validation set $D_{\text{val}}$, as a first step, we remove the so-called ``global explainers'' from the proponent and opponent list, i.e., those train examples that have a high probability of being among the top 20 proponents or opponents ~\cite{Barshan2020}.

We then extract feature maps from the last hidden layer of the network. For each feature map $k$ and for each pair of classes ($j,i$), we perform the following steps:
\begin{itemize}
\item we measure the feature map on each $z_j \in D_{\text{train}}$;
\item we take the maximum value on the obtained map, $f(z_j)_{k}$;
\item with the same procedure, we compute $f(z'_i)_{k}$ for each $z_i' \in D_{\text{test}}$;
\item we define a similarity metric between $f(z_j)_{k}$ and f$(z'_i)_{k}$ as:
\begin{equation}
   S(z_j,z'_i)_k =  \frac{f(z'_i)_{k} \cdot f(z_j)_{k}}{\mathrm{max}\left(|f(z'_i)_{k}|, |f(z_j)_{k}|\right)^2};
  \label{eq:similarity}
\end{equation}
\item for each $z_i'$, we order $S(z_j,z'_i)_k$ in descending order of $\widetilde{T}_{i,j}$  in such a way that the first $S(z_j,z'_i)_k$  is relative to the first proponent, the second to the second proponent and so on; 
\item we define the feature importance $FI$ of the feature $k$ for the class $i$ and the train image $z_j$ by averaging the ordered $S(z_j,z'_i)_k$ over the test set:
\begin{equation}
    FI(z_j)_{k,i} = \langle   S(z_j,z'_i)_k   \rangle_{D_{\text{test}}}.
    \label{eq:FI}
\end{equation}
\end{itemize}
$FI(z_j)_{k,i} $ represents the correlation between the latent representation of $z_j$ and $z'_i$ averaged over the test set. The behaviour of the scatter plot of $FI(z_j)_{k,i}$ ranked as described above with respect to $z_j$, gives a qualitative measure of the impact of the corresponding feature on the network decision. 
If the ordering of $S(z_j,z'_i)_k$ with respect to the TracIn score is faithful to the network decision model, we expect to see scatter plots of $FI(z_j)_{k,i} $ with non-random structures. In this case, we can select the most important features for the network decisions. In fact, we expect that the feature importance of important features is higher for proponents than for neutral examples or opponents.
We can verify the goodness of the selection by providing a quantitative metric which is the impact a feature has on the network performances. We call $y_{\mathrm{pred}}$ the prediction of the network obtained using
the complete set of features and $y_{\mathrm{pred}}^k$ the prediction obtained excluding the feature $k$ by masking the $k-$th filter of the last hidden layer of the UNet.
 
The difference in Dice score in the two configurations above is the impact of the feature $k$ in the prediction of class $i$, $F_{k,i}$:

\begin{equation}
    F_{k,i}= \langle D_i(y_{\mathrm{pred}}) - D_i(y_{\mathrm{pred}}^k) \rangle_{D_{\text{val}}}.
    \label{eq:kernel_inf}
\end{equation}

\begin{figure*}[!h]
  \centering
\includegraphics[width=\linewidth, scale = 0.2]{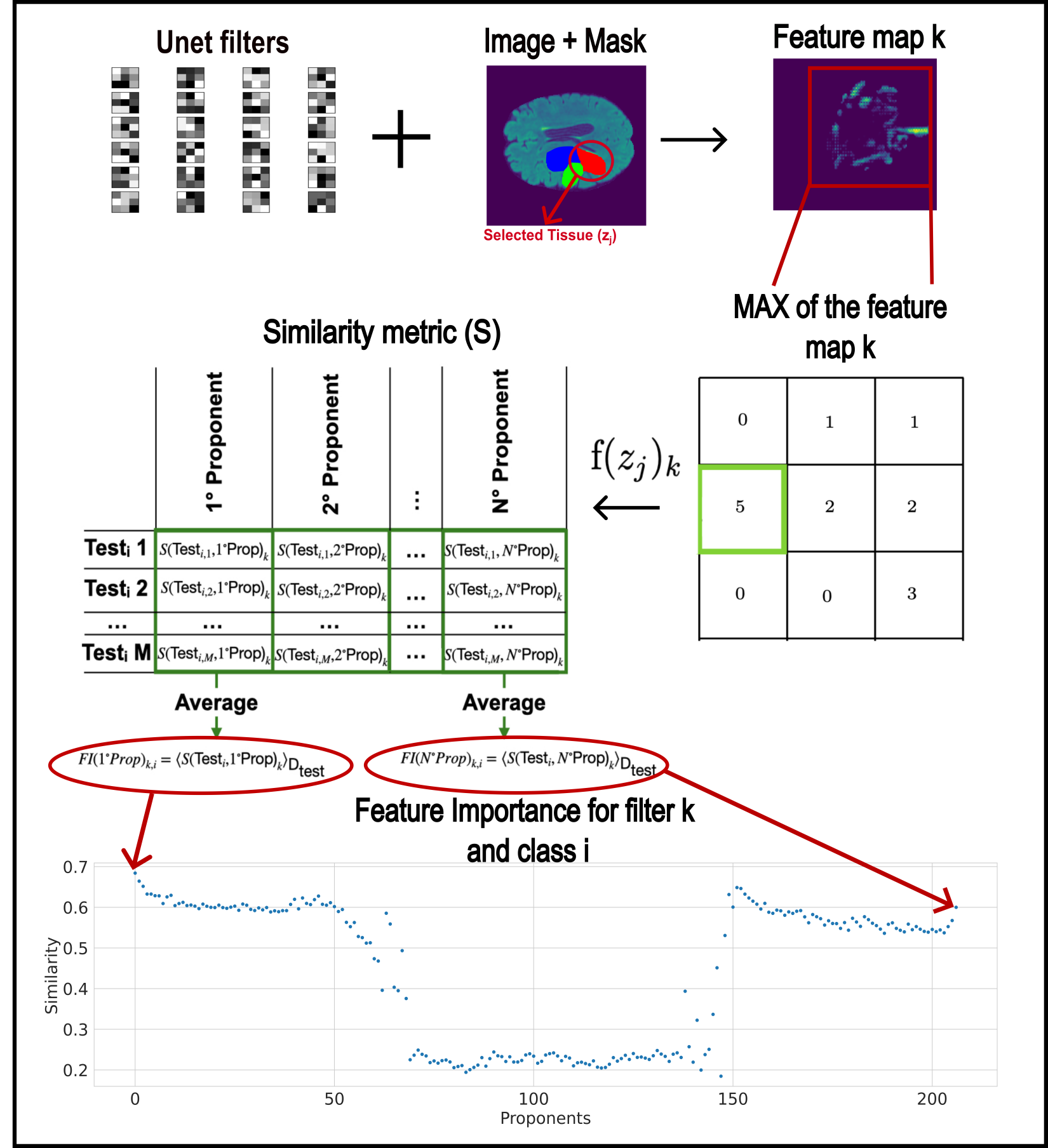}
\caption{Scheme of the feature importance calculation. For first we extract the filters from the last hidden layer of the UNet and we measure the feature maps only on the selected tissue of the images on the training and test set. We then take the maximum value on the obtained maps and we evaluate the similarity metric (S) as described in the text. Taking the average of S with respect to the test set we define the feature importance for filter $k$ and class $i$ ($FI(z_j)_{k,i}$)}
\label{fig:TFI}
\end{figure*}

Figure~\ref{fig:TFI} shows a scheme that describes the feature importance calculation.

\section{Results}
\label{Results}
\subsection{Algorithm characterization}
\begin{figure}[!h]
  \centering
  \includegraphics[width=0.75\linewidth, scale = 0.2]{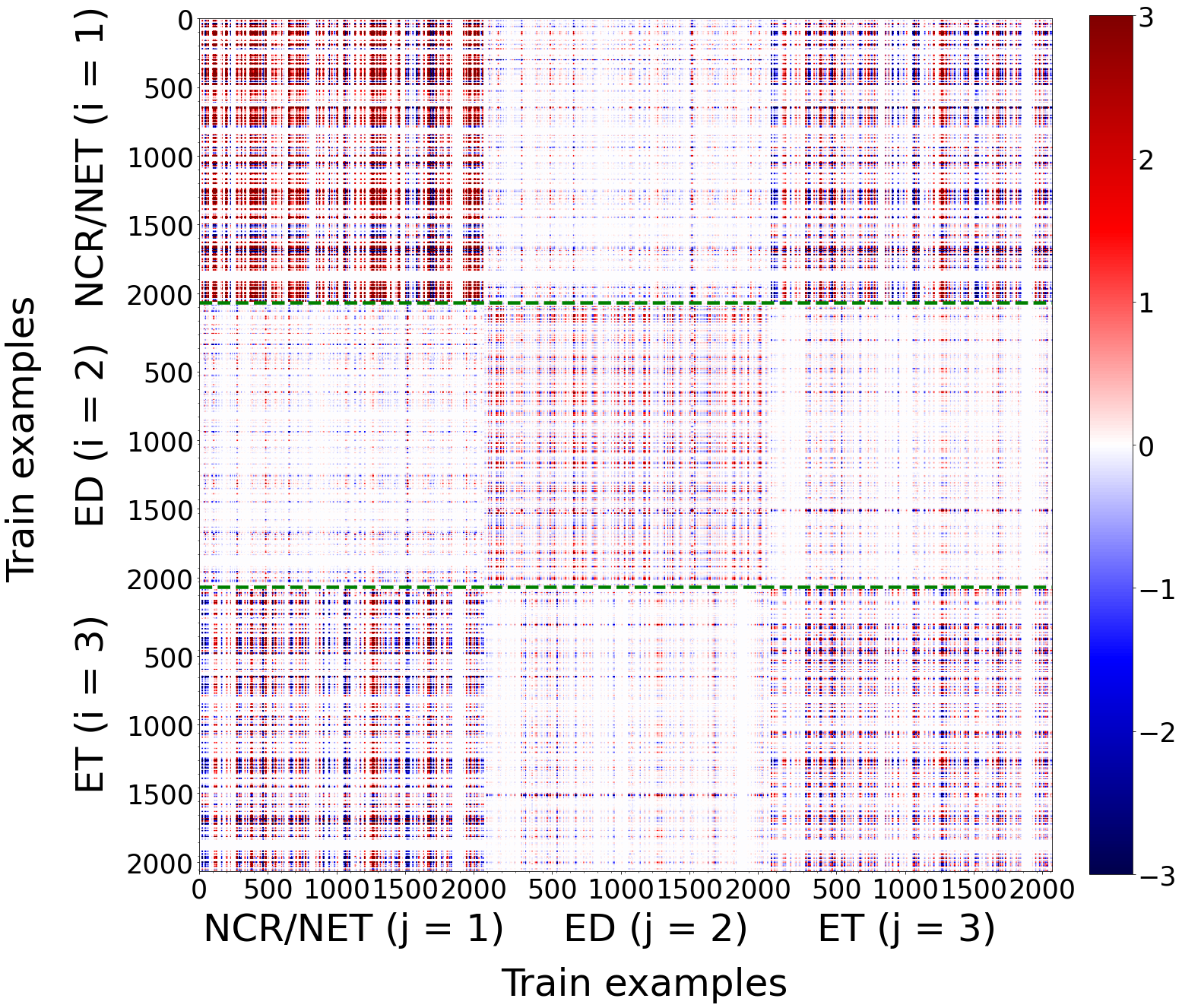}
  \caption{Train-influence matrix for the BraTS19 training set. The diagonal terms ($i$ = $j$) represent the influence that the region of $z$, predicted to be of class $i$, has on the predicted region of the same class of $z'$, while off-diagonal terms quantify the influence of regions of $z$ predicted to be of classes different from $i$ 
  }
  \label{fig:self}
\end{figure}

The three train-influence matrices are shown together in Figure \ref{fig:self}. The first row of blocks is relative to $T(z_1')$, (NCR/NET), the second to $T(z_2')$ (ED) and the third to $T(z_3')$ (ET).

For example, the first block of $2070 \times 2070$ scores represents the influence that segmented regions of class 1 have on regions predicted to be of the same class, while the second block of the first row represents the ``mixed'' influence that segmented regions of class 2 have on regions predicted to be of class 1, and so on. White rows and columns are cases where the corresponding label is not present in the slice under consideration.

It can be noted that the influences of the different blocks are well separated and that the TracIn score is quite uniform within each block, thus being not sensitive to differences among regions of slices predicted to be of the same class, but only to differences between classes. 
The absolute average value of the score of each block on the diagonal is related to the different progress in the training process that is selected by the checkpoints. For example, class 1 is the one with the lowest performances (see Table ~\ref{tab:score}) and thus has the higher value of the influences when considering regions predicted to be of the same class, due to the still high value of the gradients.

\begin{figure}[!h]
  \centering
\includegraphics[width=\linewidth, scale = 0.8]{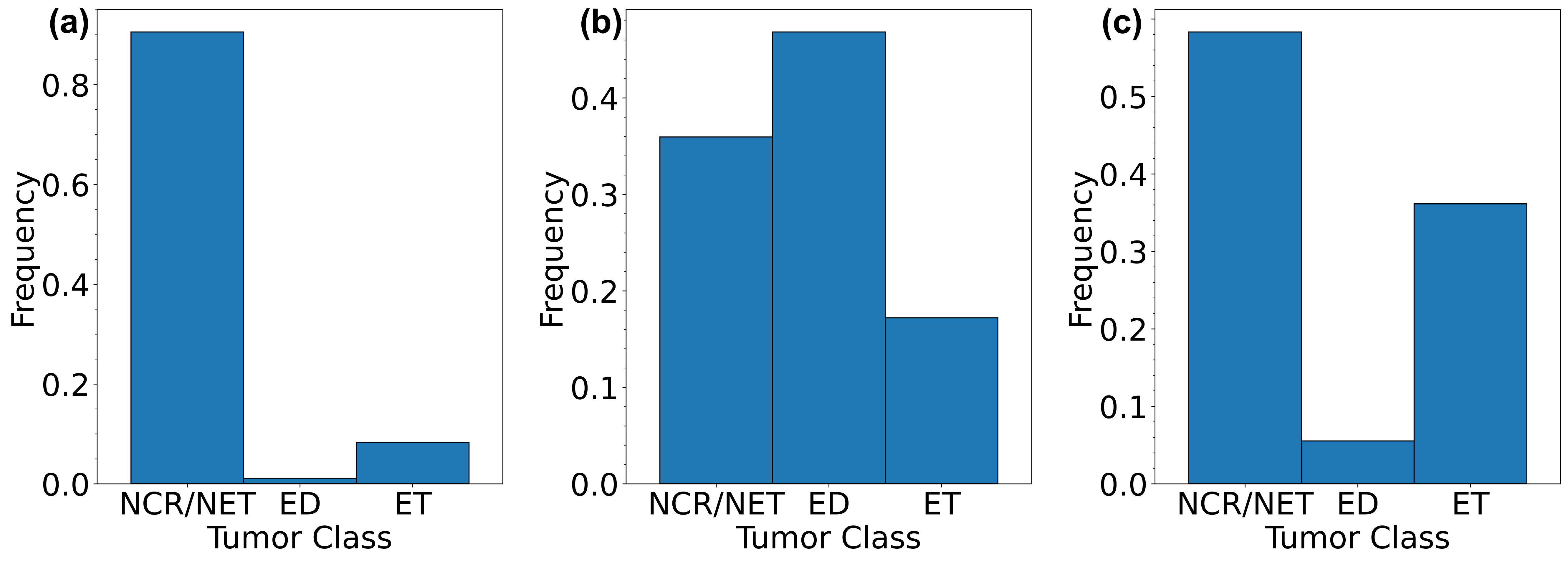}
  \caption{Distribution of the class of the top 10 proponents for the three train-influence matrices}
  \label{fig:hist}
\end{figure}

After eliminating the global explainers, as described in Section \ref{subsec:Int}, we show in Figure \ref{fig:hist} the distribution of the class of the top 10 proponents, separately for each of the three train-influence matrices.
For class 1 and 2, the top 10 proponents belong to the same class, as expected, meaning that tumors of the same class have the highest influence. This is not true for class 3, where the top 10 proponents belong more often to class 1.
Class 1 has a strong influence on all other tumor classes, which is linked to the low performances and high gradients at the checkpoint considered. On the other hand, class 3 (and similarly, class 2) have better performances than class 1 and therefore smaller gradients.

From a ``clinical'' point of view, class 3 tissue, enhancing tumor, is more similar to class 1, necrotic tumor, than to class 2, edema, which has a higher water content with respect to the others, so the two can be confused by the MRI signal.

We conclude from this study that, given our definition of TracIn, the best proponents of a given tumor in a test image ($z_i'$) is most likely a training image where a tumor of the same type is present, or with a tumor with similar MRI properties. Care must be taken when the segmentation network used provides significant differences in the segmentation performances for the different classes, since big differences in the magnitude of the gradients at the checkpoints can contaminate this trend.

\subsection{Interpretation maps and faithfulness metric}

\begin{figure*}[!h]
  \centering
  \includegraphics[width=\linewidth, scale = 0.2]{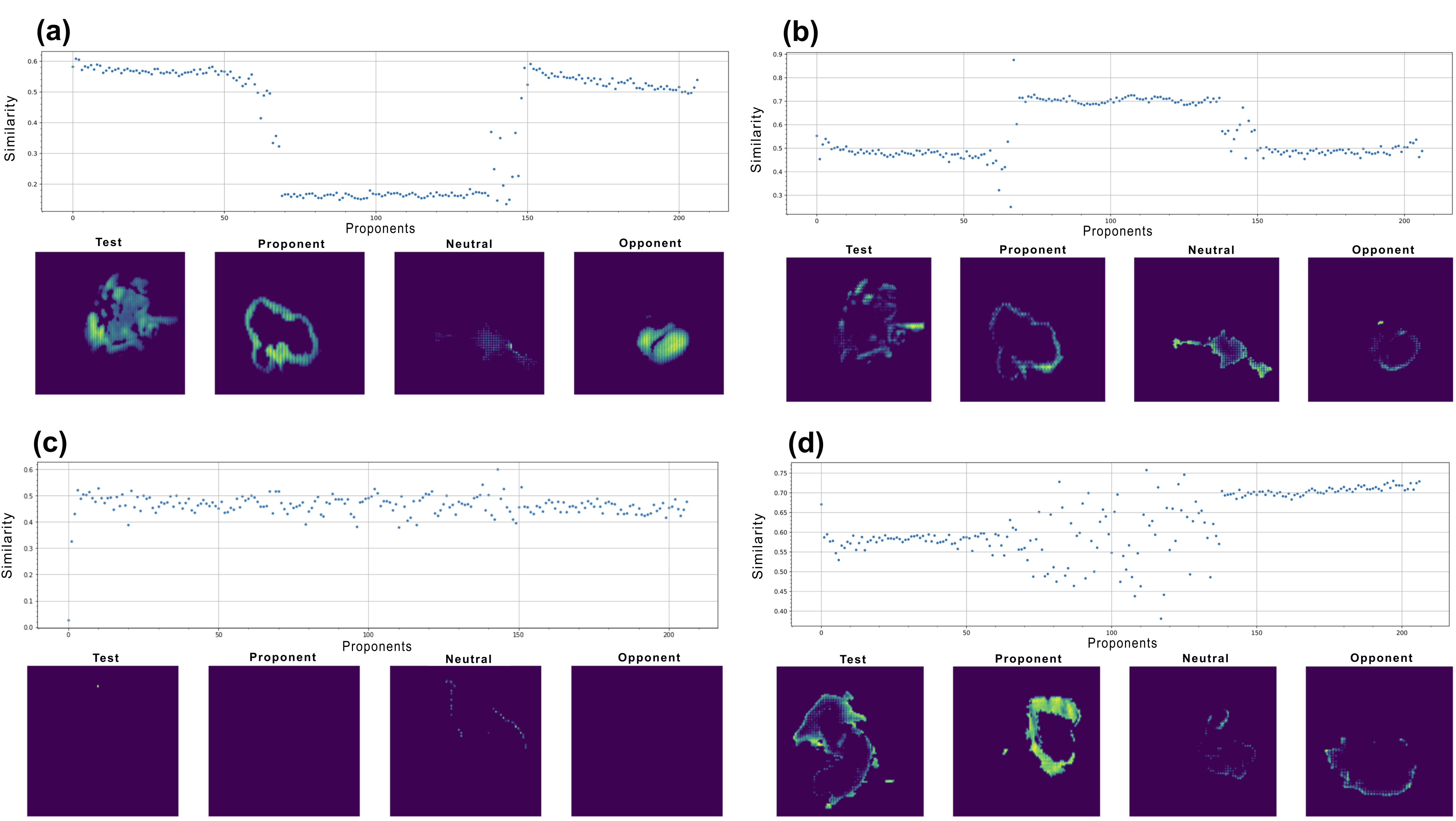}
  \caption{Examples of ranked scatter plots of $FI(z_j)_{k,i}$ evaluated using the methodology described in Section \ref{subsec:Int}. For each feature considered, we show in the upper part the ranked scatter plot, and in the lower part the features applied to images of the test and train set. In order, from left to right, a test image, one of its proponents, neutral and an opponent are shown}
  \label{fig:TFI2}
\end{figure*}

We use the 64 feature maps in the last hidden layer of the UNet, before the network prediction. Figure ~\ref{fig:TFI2} shows four ranked scatter plots of $FI(z_j)_{k,i}$ as described in Section \ref{subsec:Int} where we observe different trends.

In Figures ~\ref{fig:TFI2} a) and b) the features distinguish neutral train images (in the middle of the x-axis) from proponents and opponents. These features are used to make a coarse decision between classes.  In Figure ~\ref{fig:TFI2} c) the feature does not distinguish between train images, so we expect it is not important for the prediction and it can be removed without reducing the network performance. In Figure ~\ref{fig:TFI2} d) the plot exhibits a linear trend and the corresponding feature is used to separate proponents from opponents. If the slope is positive as in this case then the feature favours the opponents, confusing the network. Instead, in the case of a negative slope, it helps the predictions.

\begin{figure*}[!h]
    \centering
    \includegraphics[width=\linewidth, scale = 1.0]{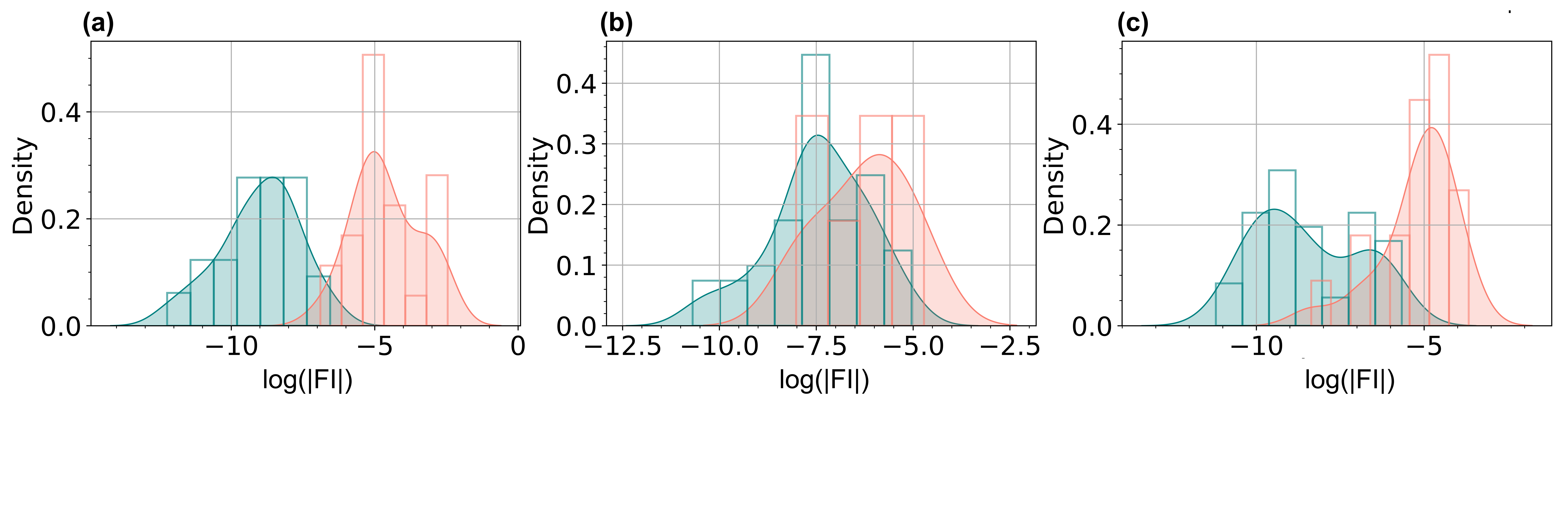}
    \caption{Impact of features (F, Equation ~\eqref{eq:kernel_inf}) on network performances in the different classes. In red are the features selected using the method described in the text, and in blue are the others. Continuous lines describe the density function associated with the histogram.}
    \label{fig:kernel_inf}
\end{figure*}

We call predictive features those with the same behaviour of Figure ~\ref{fig:TFI2} a) or with linear behaviour and a negative slope. We also ask for a feature importance (Equation ~\eqref{eq:FI}) greater than 0.65 for the first proponent. We analyze quantitatively the goodness of our predictive feature selection by computing the impact on the network performances as described in Equation ~\eqref{eq:kernel_inf}. The results are shown in Figure \ref{fig:kernel_inf} where in red are the features selected with the criteria described above and in blue the others. Since there is a significant separation, we can conclude that our TracIn score is able to effectively capture the impact that individual features have on the model prediction, thus providing a link between our explanation framework and the inner latent representation of the network. 
We are unable to associate predictive features with characteristics of the medical image that are easily recognizable to an untrained eye.

\pagebreak

\section{Conclusion and discussion}
Explainability of AI methods in medical applications is still a challenge and a barrier to their clinical use, due to the lack of transparency in the decision process of the complex models used, like deep neural networks. 

This paper proposes a methodology to implement an influence-based explainability algorithm in a medical image analysis task, where visual methods like saliency maps are often used. 

We choose the task of brain tumor segmentation in multimodal image analysis, exploiting the dataset and the resources of the BraTS19 challenge. Images in this dataset are related to gliomas, a common brain tumor, where three different types of tissues are usually present: NCR/NET (non-enhancing tumor/necrotic tumor, class 1), ED (edema, class 2), ET (enhancing tumor, class 3). Each patient MRI is composed of several slices in the sagittal direction taken in four modalities: T1-weighted, T2-weighted, T2-Flair, and post-contrast T2-weighted. Ground truth is provided by experienced radiologists.
The AI model used to perform the task is a standard 2D UNet, often used in this kind of application, with a Dice-based loss, reaching classification performances comparable to those of the challenge.

We extend the TracIn method, recently proposed and used mainly for classification tasks. For each instance predicted by a network, TracIn produces an explanation, which is a score assigned to each example of the network training set. The higher the score, the higher the influence the train example had in the prediction. According to the TracIn score, the training set can be divided into proponents, neutral and opponents examples. In our specific case, the input of the network are images (slices) of the patient's MRI, where the modality is treated as a color channel. 


Compared to the classification task, segmentation is more complex, since it involves the classification of several pixels together. As a consequence, influence-based explainability methods can be spoiled by the contribution of heterogenous pixels and are more prone to ``noise'' effects.

To limit the impact of the effect above, we define a TracIn score separately for each of the tumor classes. Moreover, by applying a threshold on the network output, we reduce the contribution of heterogenous pixels.

One possibility is to use TracIn to increase the trust a radiologist has in the tumor segmentation performed by the network, by providing a local, example-based explanation, i.e. showing a list of proponents and opponents that give insights on the decision process.
To investigate this possibility, we studied the train-influence matrices (i.e. the influence of the training set on itself), and we observed that the strongest proponents of an image with a tumor of a given type (class) are more often train images with the same type of tumor, while the explanation is not sensitive to differences in tumor of the same type.  This particular conclusion is relative to our specific use case (gliomas), while we cannot exclude that in other cases there is sensitivity among tumors of the same types.

We also investigated the use of our TracIn method to provide a global explanation of the segmentation network. In particular, we evaluated the faithfulness of the explanation with respect to the network decision by using a feature importance metric that links the output of the TracIn with the latent representation of the network. 

 This metric can be applied to evaluate features' importance and to guide the user's interpretation of the explanation. The goodness of our selection of the important features is confirmed by the analysis of the impact on the network performances. We then extend TracIn explanation not only by producing a list of proponents and opponents without context but also by giving the user information about which features were used to make the decision. The fact that the TracIn score is directly correlated with the latent representation of the network proves the faithfulness of the explanation to the network decision model. By using this xAI method and correlating it with the extracted features, we can identify not only which examples of the train were used by the network to aid prediction, but also why (based on which features) and how (how the features are used to make the decision).
\newline
\newline
\newline
\noindent \textbf{Acknowledgement} This work was funded by the European Union’s CHIST-ERA programme under grant agreement CHIST-ERA-19-XAI-009 (Call Topic: Explainable Machine Learning-based Artificial Intelligence (XAI), Call 2019), and supported by PNRR MUR project PE0000013-FAIR.
\newline
\noindent \textbf{Statements and Declarations} The authors declare no conflict of interest



\end{document}